\documentclass[aps,pre,superscriptaddress,%
 amsmath,amssymb,
reprint,%
]{revtex4-2}
\usepackage{float}
\usepackage{graphicx}
\usepackage{dcolumn}
\usepackage{bm}

\usepackage[T1]{fontenc}
\usepackage[english]{babel}
\usepackage{mathptmx}
\usepackage{xcolor}
\usepackage{booktabs}

\usepackage{pgfplots}
\pgfplotsset{compat=1.15}
\usepackage{tikz}
\usetikzlibrary{decorations.pathreplacing, arrows.meta}

\newcommand{\be}{\begin{equation}}
\newcommand{\ee}{\end{equation}}
\newcommand{\bea}{\begin{eqnarray}}
\newcommand{\eea}{\end{eqnarray}}

\usetikzlibrary{decorations.markings}
\newcommand*{\A}[1]{\fill[red]   #1 ++(0,-.12) -- ++(.35,0) -- ++(.1,.12) --
                                  ++(-.1,.12) -- ++(-.35,0) -- cycle;}
\newcommand*{\T}[1]{\fill[blue]  #1 ++(0,-.12) -- ++(.35,0) -- ++(-.1,.12) --
                                  ++(.1,.12) -- ++(-.35,0) -- cycle;}
\newcommand*{\G}[1]{\fill[green] #1 ++(0,-.12) -- ++(.35,0) arc(-90:90:.12)
                                  -- ++(-.35,0) -- cycle;}
\newcommand*{\C}[1]{\fill[yellow]#1 ++(0,-.12) -- ++(.35,0) arc(270:90:.12)
                                  -- ++(-.35,0) -- cycle;}

\newcommand{\BasePair}[3]{%
  \begin{scope}[shift={#1},rotate=#2]
    \ifcase#3\relax
    \or 
      \A{(0,0.22)} \T{(0,-0.22)}
    \or 
      \T{(0,0.22)} \A{(0,-0.22)}
    \or 
      \G{(0,0.22)} \C{(0,-0.22)}
    \or 
      \C{(0,0.22)} \G{(0,-0.22)}
    \fi
    \draw[dashed,gray] (0,0.05) -- (0,-0.05);
  \end{scope}
}

\usetikzlibrary{decorations.markings}
\begin{document}


\title[]{Synthetic design of force-responsive hydrogels with ring-forming catch bonds}

\author{Wout Laeremans}
\affiliation{Soft Matter and Biological Physics, Department of Applied Physics and Science Education, and Institute for Complex Molecular Systems,
Eindhoven University of Technology, P.O. Box 513, 5600 MB Eindhoven, Netherlands}
\author{Wouter G. Ellenbroek}
\email{w.g.ellenbroek@tue.nl}
\affiliation{Soft Matter and Biological Physics, Department of Applied Physics and Science Education, and Institute for Complex Molecular Systems,
Eindhoven University of Technology, P.O. Box 513, 5600 MB Eindhoven, Netherlands}

\date{\today}

\begin{abstract}
Catch bonds are interactions whose lifetimes increase under mechanical load, a counterintuitive behaviour that underlies diverse biological processes. Translating this mechanism to synthetic materials offers the potential to create systems that are compliant at low stress but stiffen under applied force, with applications ranging from impact-responsive materials to dynamic tissue scaffolds. However, engineering materials with tunable, force-dependent interactions remains challenging, and existing conceptual designs are limited. Here, we present a minimal synthetic framework for catch bond behaviour in dynamic hydrogels, based on reversible ring-forming polymers. Using coarse-grained molecular dynamics simulations, we show that hydrogels with such a chemistry undergo fewer bond-breaking reactions as the stress increases and can even display a non-monotonic dependence of the strain rate on the applied stress. Our results highlight the potential of reversible ring formation as a versatile platform for designing mechanically adaptive materials with tunable durability and responsiveness.
\end{abstract}

\maketitle


In most systems, whether biological or engineered, the lifetime of a bond decreases exponentially with increasing mechanical load, as described by classical models of force-induced rupture~\cite{bell1978models, evans2001probing}. Intriguingly, a special class of interactions defies this general rule: catch bonds strengthen under tension, exhibiting prolonged lifetimes up to a characteristic force before eventually weakening~\cite{dembo1988reaction}. Over the past two decades, this counterintuitive behaviour has been observed in numerous receptor–ligand pairs across a wide range of biological contexts~\cite{milles2018molecular, mathelie2020force, marshall2003direct, sauer2016catch, buckley2014minimal, litvinov2018regulatory, melani2022blood, lim2017single, liu2020high, harder2015catch, yago2008platelet, liu2014accumulation}. Furthermore, several theoretical frameworks have been developed to rationalize these phenomena and to extract quantitative parameters from experimental data~\cite{thomas2008biophysics, thomas2006catch, pereverzev2009allosteric, lou2007structure}. Yet, despite these advances, such models remain predominantly phenomenological. The intricate and multidimensional energy landscapes governing biological catch bonds continue to challenge a complete mechanistic understanding of how mechanical forces can stabilize molecular interactions.

In materials science, catch bonds represent an attractive design principle due to their force-dependent strengthening behaviour~\cite{dansuk2023catch, rabbi2025fine}. Such interactions enable materials to remain compliant under low mechanical loads while resisting deformation at higher stresses~\cite{dansuk2019simple}. This property has potential applications ranging from impact-responsive systems like protective gear~\cite{dansuk2023catch} to cell-interactive scaffolds that are initially soft to permit cell migration but stiffen under contractile forces~\cite{ooi2017hydrogels}. Moreover, catch bonds enable active network remodeling under load, whereby transient unbinding allows bonds to redistribute toward high-stress regions. This can suppress crack initiation and enhances overall mechanical robustness~\cite{mulla2022weak}. To engineer materials that exhibit catch bond behaviour, a synthetic design capable of reproducing this force-dependent interaction is required, a task that remains far from trivial~\cite{van2020chemical}. Several simple design concepts have been proposed in recent years~\cite{dansuk2019simple, dansuk2021self, dansuk2023catch, van2020chemical, liu2024engineering, yang2024engineering}, but their number remains limited. In this spirit, we propose a new design based on reversible ring-forming polymers.

With reversible ring formation, we refer to a reaction that occurs when two reactive sites along a linear polymer chain come into proximity, resulting in the formation of a cyclic fragment and the cleavage of the remaining chain. Such chemistries have only recently been established experimentally as a route to covalently adaptable polymer networks~\cite{van2021reprocessing, delahaye2019internal}. To date, these materials have been primarily investigated for their reprocessability and recyclability~\cite{van2021reprocessing}. Beyond this established role, reversible ring formation also introduces a fundamentally different and largely unexplored possibility: force-dependent stabilization of network connectivity reminiscent of catch bond mechanics. In this light, reversible ring formation represents a conceptually simple synthetic design for catch bonds, with particular promise for multifunctional hydrogels that combine reprocessability with stress-responsive mechanical adaptation.

In this work, we investigate the mechanical potential of materials incorporating reversible ring-forming reactions using molecular dynamics simulations. To probe their adaptive response, we subject these networks to tensile tests and examine how the material evolves under stress. In particular, we show that the number of reactions reduces under increasing load and how these microscopic dynamics translate into a non-monotonic macroscopic relation between stress and strain rate. These results provide theoretical evidence that polymers capable of reversible ring formation can generate adaptive mechanical responses, where microscopic topology changes produce emergent, tunable macroscopic properties. This establishes such systems as promising candidates for the design of multifunctional, stress-responsive hydrogel materials.


\section*{Results}
\label{sec:catchbond_mech}
\subsection*{Catch bond design}
\begin{figure}[t]
    \centering
    \includegraphics[width=\linewidth]{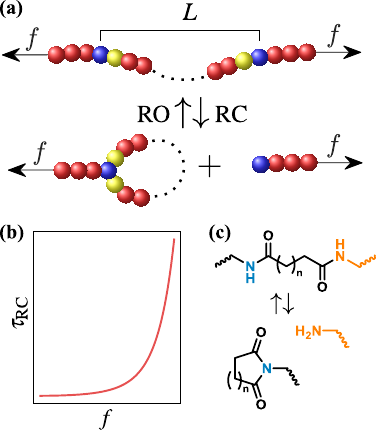}
    \caption{(a) Coarse-grained visualisation of a reversible ring-forming polymer. Dotted lines schematically indicate that the connected sites may be separated by an arbitrary number of beads in the coarse-grained model. Going from the bottom to the top configuration, is referred to as ring-opening (RO), while going from the top to the bottom configuration is referred to as ring-closing (RC). The reaction occurs when the reactive (blue) groups are in close proximity. (b) The higher the force $f$, the longer it takes for these reactive groups to encounter each other, thereby increasing the average lifetime $\tau_\text{RC}$ before ring-closing. Hence, this system represents a catch bond. At very high forces, bonds between individual atoms would break, but this is outside the regime of interest in this work. (c) An example of reversible ring formation in the form of the bisamide exchange reaction~\cite{van2021reprocessing}. Here, a polymer strand contains two amide groups, which can react to form a cyclic imide and a free amine.}
    \label{fig:Fig1}
\end{figure}

Reversible ring-forming polymers are macromolecules capable of dynamically interconverting between linear chains and cyclic fragments via intramolecular reactions that form rings and cleave off chain segments. At a coarse-grained level, this system can be visualised as in Fig.~\ref{fig:Fig1}(a), where the blue segments represent reactive groups that undergo a reversible reaction upon close proximity. When a tensile force is applied to the chain ends, these reactive groups encounter each other less frequently, leading to an increase in the average lifetime $\tau_\text{RC}$ of the intact linear configuration, as illustrated in Fig.~\ref{fig:Fig1}(b). Crucially, ring formation converts a continuous, load-bearing chain into two shorter segments, such that the suppression of this reaction under tension effectively prolongs the mechanical integrity of the chain, consistent with catch bond behaviour. Such chemistries have been realised experimentally as well~\cite{van2021reprocessing, delahaye2019internal}. One example consists of the bisamide exchange reaction~\cite{van2021reprocessing}, as visualised in Fig.~\ref{fig:Fig1}(c). Here, a polymer strand contains two amide groups, which can react to form a cyclic imide and a free amine.

At the single chain or microscopic level, several theoretical studies have investigated the effect of tension on the ring formation time~\cite{blumberg2005disruption, shin2012effects, laeremans2024polymer, laeremans2025theoretical}. For example, when the polymer is approximated as a freely jointed chain, the influence of an applied force can be captured by the simple relation~\cite{laeremans2024polymer}
\begin{align}
\tau_\text{RC}(L;f) \sim L^{3/2} \left[ \frac{\sinh(\beta f b)}{\beta f b} \right]^{L}, \label{eq:scaling}
\end{align}
with $L$ the length in between the reactive groups, $\beta = (k_\mathrm{B}T)^{-1}$ the inverse temperature and $b$ the bond length. In more complex cases, where the chain bending stiffness becomes significant, the free energy landscape must be considered explicitly~\cite{blumberg2005disruption, shin2012effects, laeremans2025theoretical}. At very high forces ($\gtrsim 5\,\mathrm{nN}$), covalent bonds between atoms may rupture, independent of ring formation. However, such forces are far beyond the range of interest in this work. By contrast, ring formation is already significantly suppressed at forces on the order of $1\,\mathrm{pN}$, which corresponds to the typical entropic force scale governing polymer conformations.

Thus, ring-forming polymers offer a conceptually straightforward framework for designing catch bonds, using chemistries that are experimentally accessible. In these systems, ring formation acts as a cleavage reaction that breaks load-bearing polymers, but this reaction is suppressed under increasing force, giving rise to force-dependent stabilization. Although this catch bond behaviour may appear straightforward at the single chain level, predicting how it translates to a polymeric network or bulk material composed of reversible ring-forming polymers remains nontrivial. Specifically, one may ask whether a macroscopic system built from these molecular units will also exhibit catch bond behaviour. This question is addressed in the remainder of this work.

\subsection*{Macroscopic response under deformation}
To investigate the macroscopic response, we perform a volume-preserving tensile test: the polymer network is stretched in one direction while contracted in the perpendicular directions, by imposing a target stress along the stretching axis. Given the initial dimensions $\ell \times \ell \times \ell$, we apply a target stress, resulting in a time-dependent deformation of the form
\begin{align}
    \mathbf{F}(t) &= \text{diag}(\lambda(t), 1/\sqrt{\lambda(t)}, 1/\sqrt{\lambda(t)}) ,\label{eq:deform}
\end{align}
with $\det(\mathbf{F}) = 1$, ensuring volume preservation. For a regular (non-catch bond) dynamic polymer network, the reaction rates are typically determined by the concentration of reactive groups, so no change is expected under such deformation in a homogeneous system. However, for our proposed catch bond system, we expect an effect, as explained below.

Although we simulate the full dynamics, including non-affine relaxation, for simplicity of the argument we now approximate all chains as deforming affinely. In that case, a chain with initial end-to-end vector $\mathbf{R}$ transforms under the proposed deformation to $\mathbf{R}' = \mathbf{F}\mathbf{R}$, such that 
\begin{align}
    R'^2 = \mathbf{R} \cdot \mathbf{F}^{\top} \mathbf{F} \mathbf{R}.
\end{align}
Denoting the angle between $\mathbf{R}$ and the stretching direction by $\theta$, we have (see Supplementary Material)
\begin{align}
    \frac{R'^2}{R^2} = \lambda^2 \cos^2(\theta) + \frac{1}{\lambda} \sin^2(\theta).
\end{align}
Averaging over the solid angle and assuming homogeneous chain orientations before stretching ($\langle \cos^2(\theta) \rangle = 1/3$, $\langle \sin^2(\theta) \rangle = 2/3$) yields~\cite{rubinstein}
\begin{align}
    \langle R'^2 \rangle &= \frac{\lambda^2 + 2/\lambda}{3} \, \langle R^2 \rangle. \label{eq:Raffine}
\end{align}
Thus, although chains oriented primarily perpendicular to the stretching direction shorten in end-to-end distance, the chains are, on average, still stretched. As stretching of a chain increases the lifetime before ring formation (see Fig.~\ref{fig:Fig1}), the number of chain splitting reactions within the network is expected to lower under such tensile tests. Consequently, the suppression of ring formation under tension is expected to stabilize the network by preserving load-bearing chains, effectively prolonging the material’s mechanical integrity under applied stress. In the Supplementary Material, we furthermore show that this is not only expected for uniaxial deformations as depicted in Eq.~\ref{eq:deform}, but for any affine, volume-preserving deformation, e.g. a shear deformation.

On the macroscopic scale, the relevant quantity is the strain rate. Let $\ell_x(t)$ denote the network length along the stretching direction at time $t$. The strain is defined as
\begin{align}
    \epsilon(t) &= \frac{\ell_x(t) - \ell_x(0)}{\ell_x(0)},
\end{align}
and the corresponding strain rate is
\begin{align}
    \frac{d\epsilon(t)}{dt} &= \frac{1}{\ell_x(0)} \frac{d\ell_x(t)}{dt}.
\end{align}
In the simulations considered here, a fixed target stress is applied to the network, causing it to elongate over time. The resulting time-dependent strain is thus a creep response, where the rate of elongation reflects the underlying network dynamics. For conventional materials, the strain rate is expected to increase monotonically with applied stress, reflecting a direct correlation between applied force and deformation rate. In this work, we show that polymer networks containing reversible ring-forming catch bonds can instead exhibit a non-monotonic relationship between strain rate and applied stress. At low stresses, the network extends slowly, while at high stresses it elongates more rapidly. However, at intermediate stresses, the response is reversed: increasing the applied stress leads to a lower strain rate due to the force-dependent dynamics of the catch bonds. This behaviour constitutes a macroscopic manifestation of catch bond mechanics.

\begin{figure}[t]
    \centering
    \includegraphics[width=\linewidth]{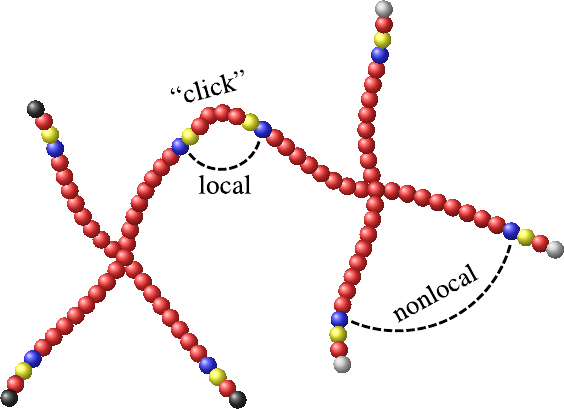}
    \caption{To mimic a click chemistry, star polymers with two different end groups (black / grey) are placed in a simulation box. By diffusion, when the ends are in close proximity, they irreversibly bind or ``click''. After the gelation process, the reversible ring-forming reaction is turned on (see Fig.~\ref{fig:Fig1}). A distinction is made between local and nonlocal ring-closing. A local reaction refers to a reaction initiated by two reactive groups on the same chain, which in our model corresponds to groups separated by exactly six beads, while a nonlocal reaction occurs between any other pair of reactive groups.}
    \label{fig:Fig2}
\end{figure}


\subsection*{Hydrogel formation by click chemistry}
To investigate the behaviour of reversible ring-forming polymers within a network, we construct a hydrogel via a gelation process that mimics click chemistry~\cite{xi2014click, oshima2014model}. This approach is a well established method for simulating polymer networks~\cite{raffaelli2021stress}. Specifically, two types of star polymers with different functionalization are employed: one bearing reactive end groups of type $A$ and the other of type $B$. When end groups of different type come into close proximity, they react to form irreversible ``click'' bonds, as illustrated in Fig.~\ref{fig:Fig2}. By introducing a large number of such stars into a simulation box with periodic boundary conditions, a percolating network emerges naturally through diffusion-driven encounters~\cite{raffaelli2021stress}. 

In this work, the stars are modeled as bead-spring polymers, in which bonded beads are connected by harmonic springs described by the potential
\begin{equation}
U_{\rm bond}(\mathbf{r}_{i}, \mathbf{r}_{i+1}) = \frac{K}{2} \left(|\mathbf{r}_{i+1} - \mathbf{r}_i| - b\right)^2,
\end{equation}
where $\mathbf{r}_j$ denotes the position of bead $j$, $b = 1$ is the rest length, and $K = 100$ is the bond stiffness, reported in the dimensionless Lennard-Jones unit system. Non-bonded beads interact via the purely repulsive Weeks-Chandler-Andersen (WCA) potential~\cite{weeks1971role}
\begin{equation}
U(r_{ij}) =
\begin{cases}
4\varepsilon \left[ \left( \dfrac{\sigma}{r_{ij}} \right)^{12}
- \left( \dfrac{\sigma}{r_{ij}} \right)^{6} \right] + \varepsilon, & r_{ij} \leq 2^{1/6}\sigma, \\[8pt]
0, & r_{ij} > 2^{1/6}\sigma,
\end{cases}
\label{eq:WCA}
\end{equation}
where $\varepsilon = 1$, $\sigma = 1.3$, and $r_{ij}$ denotes the distance between beads $i$ and $j$.

In a simulation box with periodic boundary conditions, 75 star polymers of each end-functionalization were placed at random positions. Each star consists of four arms, with each arm containing 26 beads (excluding the central bead). Consequently, each star contains a total of $26 \times 4 + 1 = 105$ beads. The reactive groups are placed on the fourth-to-last bead of each arm, such that when two ends click together, six beads separate the bonded reactive groups, as illustrated in Fig.~\ref{fig:Fig2}. The clicking mechanism is activated when two reactive end groups with complementary functionalization approach within a distance $\sigma$, corresponding to the point of contact.

\begin{figure}[t]
    \centering
    \includegraphics[width=\linewidth]{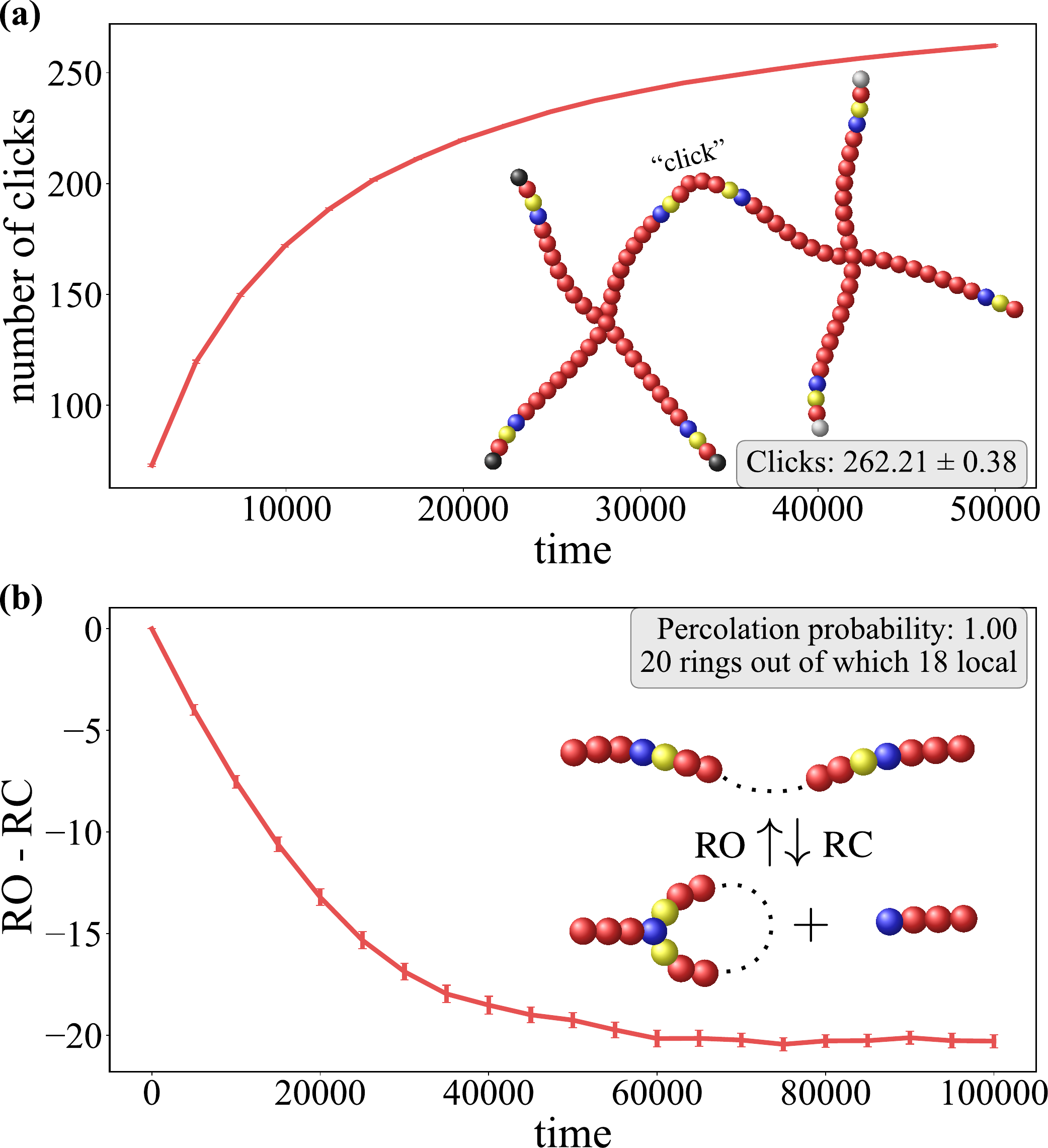}
    \caption{(a) Formation of the network by click chemistry. At the end of the simulation, on average there are $262.21 \pm 0.38$ out of 300 possible crosslinks or clicks formed. (b) Relaxation of the network with reversible reactions. On average, 20 rings are formed after relaxation, out of which 18 are local. After the relaxation protocol, unit probability for percolation in all spatial directions was observed. }
    \label{fig:Fig3}
\end{figure}

We define the bead concentration as~\cite{raffaelli2021stress}
\begin{align}
    C &= \frac{N_{\text{beads}} \sigma^3}{V},
\end{align}
where $V$ is the volume of the simulation box. We choose a concentration of $C = 7 \, \%$. Under these conditions, we observe an average of $262.21 \pm 0.38$ click bonds formed at the end of the simulation as shown in Fig.~\ref{fig:Fig3}(a), corresponding to more than $87\%$ of the maximum possible number of clicks ($75 \times 4 = 300$). The reported average is taken over 100 independent realizations, and the error denotes the standard error of the mean. Network percolation was observed in all simulations and in all spatial dimensions considered.

\subsection*{Network relaxation with reversible reaction}
After the network formation step, the click chemistry is deactivated, and the reversible reaction depicted in Fig.~\ref{fig:Fig1} is activated. However, a distinction will be made between ``local'' and ``nonlocal'' reactions. Ring-closing will be called local, when the reactive groups that initiate the reaction are located on the same chain. Looking at Fig.~\ref{fig:Fig2}, this is when two reactive groups are linked together by exactly 6 beads, 2 yellow and 4 red ones, due to the click chemistry. When other reactive groups initiate the reaction, the reaction is called nonlocal. One can chemically think of the distinction between local and nonlocal as differentiating between small-membered rings and macrocyclic systems, which both exist~\cite{ibrahim2024chemical}. The ring-opening reaction is always nonlocal, as the chain was already disconnected.

At each timestep, the distances between reactive groups are determined in order to assess whether a reaction should be initiated. For the ring-opening reaction that restores the chain, a natural choice for the reaction distance is $\sigma$, which corresponds to the characteristic size of a bead. Consequently, the sole criterion for reformation is direct contact between the reactive groups.

The choice of reaction distance for the ring-closing or cleavage reaction is largely dictated by the requirement that this reaction should be slow enough that the material does not disintegrate completely. This demand, that the network should remain gel-like at all times and not flow like a liquid, would also apply in experimental settings. To this end, we choose a reduced reaction distance. Additionally, we add a probabilistic step to the cleavage reaction to reduce the reaction rate even further.
Additional implementation details are provided in the Supplementary Material.

With the reversible reactions activated, the network initially undergoes slight degradation, which is unavoidable because the starting condition is one in which all chains are intact. Over time, however, the system reaches an equilibrium, producing a stable dynamic polymer network, as shown in Fig.~\ref{fig:Fig3}(b). For our system, we find in the equilibrium network on average 20 rings, out of which 18 are 6-membered (local). Despite the presence of reversible bond rearrangements, the network exhibits a percolation probability of unity, as expected for a mechanically stable material. We are thus able to simulate a dynamic polymer network that closely mirrors experimentally realizable systems, combining bond reversibility with long-term structural integrity.

\begin{figure}[t!]
    \centering
    \includegraphics[width=\linewidth]{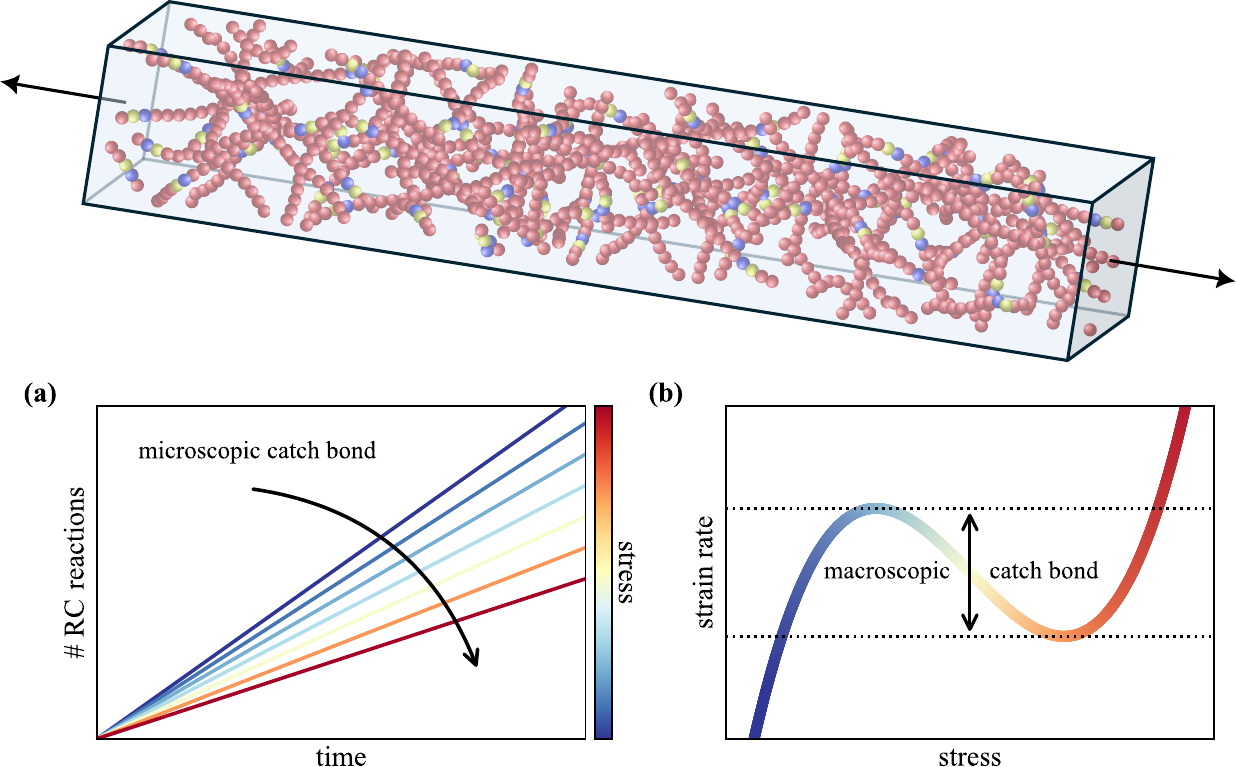}
    \caption{Tensile test illustration with two catch bond manifestations: (a) shows the microscopic manifestation, with fewer reactions occurring with increasing stress, while (b) shows the macroscopic manifestation, where the strain rate exhibits a non-monotonic dependence on applied stress.
}
    \label{fig:Figtenstest}
\end{figure}

\subsection*{Tensile tests}
The network is now relaxed and in a well-defined non-deformed starting configuration and will now be subject to the longitudinal stress discussed above. In order to be able to disentangle the deformations that a non-dynamic polymer network would have from the deformations that result from the exchange reactions, we split our numerical deformation test in two phases. In the first phase, the exchange reactions are turned off.
A constant uniaxial target stress is then applied, causing the material to deform along this axis, without any topology change.
As the deformation is volume-preserving, the other two spatial dimensions contract, as described previously in Eq.~\ref{eq:deform}. After a sufficient amount of time, the network reaches its equilibrium strain under the applied stress. Following this equilibration period, the second phase starts in which the reversible ring-forming reaction is reactivated. This induces further elongation of the pre-stressed network in the form of creep, now driven by chemical reactions rather than by the initial network topology. This splitting up of the test in two phases is a reasonable approximation of what happens in a real creep experiment, because also there the initial deformation of the network is quite sudden and the effects of the reactions happen at a much longer timescale.

As discussed previously, this protocol provides a means to probe catch bond behaviour, as visualized in Fig.~\ref{fig:Figtenstest}. In conventional materials, the number of reactions is expected to depend primarily on the concentration, which is held constant here. However, because higher applied stresses lead to more largely extended polymers (Eq.~\ref{eq:Raffine}), fewer reactions are anticipated at higher stress levels, as shown in Fig.~\ref{fig:Figtenstest}(a). Moreover, whereas typical materials exhibit higher strain rates under higher applied stresses, in the present system the opposite trend may emerge at intermediate stress levels, owing to the reduced number of reactions occurring, as depicted in Fig.~\ref{fig:Figtenstest}(b). At low stress, the network extends slowly, and at high stress, it elongates rapidly. But at intermediate stresses, the trend reverses: higher stress lowers the strain rate due to the force-dependent dynamics of catch bonds, revealing a macroscopic signature of catch bond mechanics.

\subsection*{Reduced number of reactions}
\begin{figure}[t!]
    \centering
    \includegraphics[width=\linewidth]{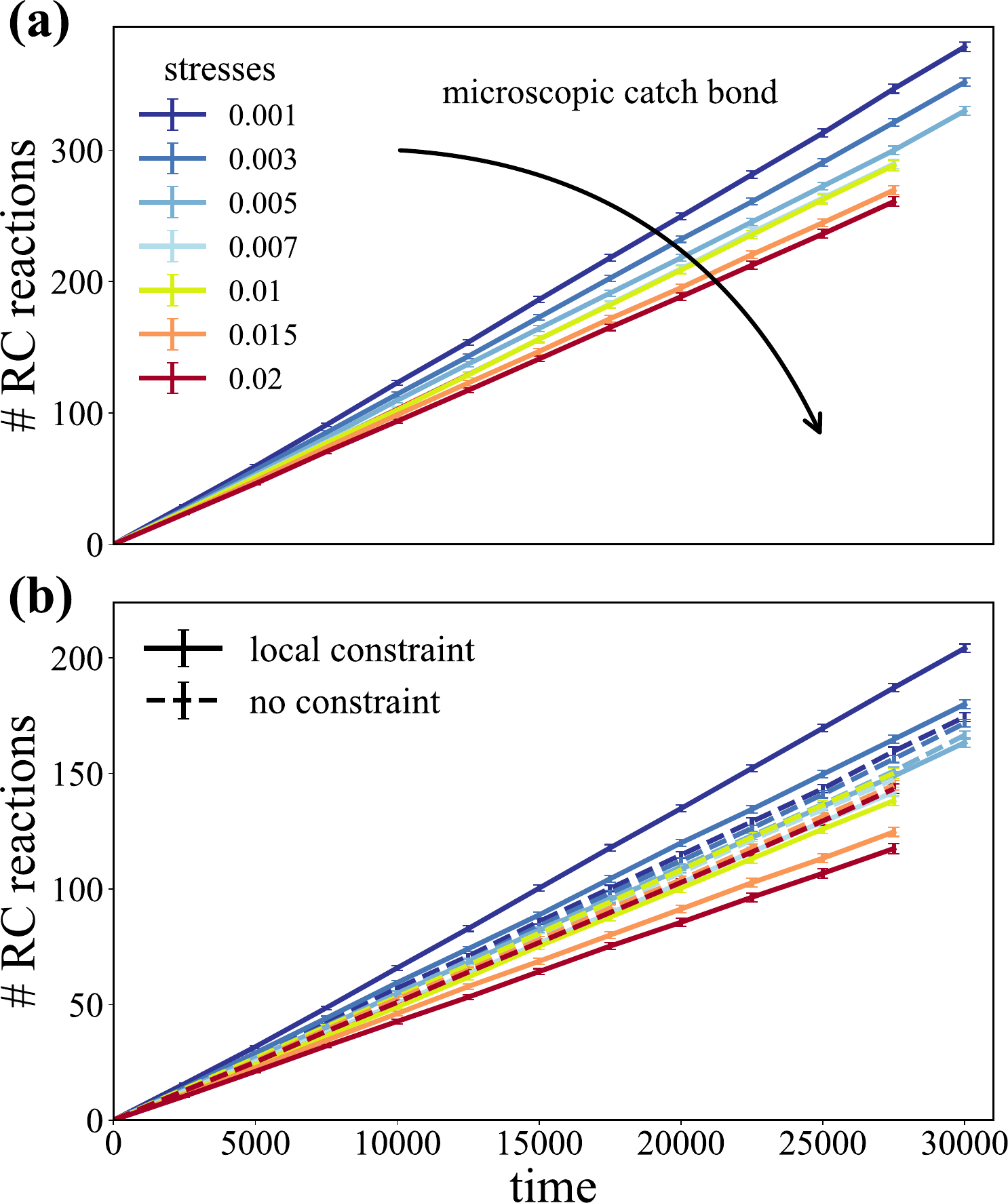}
    \caption{Microscopic catch bond behaviour under applied uniaxial stress. (a) Number of RC reactions as a function of time during a tensile test for fixed target uniaxial stress. Increasing the stress leads to a reduced number of reactions, showing a catch bond effect. (b) Again the number of RC reactions, but now distinguishing between strictly local reactions (solid lines), and reactions without the locality constrained (dotted lines). The stress dependence is markedly stronger for strictly local reactions, indicating that the microscopic catch bond effect predominantly originates from local reactions.}
    \label{fig:FigResMicro}
\end{figure}

The number of cleavage (ring-closing) reactions was measured during the tensile test, as presented in Fig.~\ref{fig:FigResMicro}(a). A clear trend emerges: increasing the applied stress leads to a reduced number of reactions, despite the concentration of reactive groups being constant. According to Eq.~\ref{eq:Raffine}, higher stresses result in greater average stretching of the polymer chains, which increases the lifetime of a chain segment before ring formation and therefore suppresses reactions. While this effect is well understood at the level of a single chain~\cite{blumberg2005disruption, shin2012effects, laeremans2024polymer, laeremans2025theoretical}, our results demonstrate that the same reasoning remains valid in a crosslinked polymer network. As ring-closing reactions split the polymer chain into two fragments, they temporarily reduce the number of load-bearing chains. As a result, suppressing these reactions under an applied stress preserves chain continuity, thereby increases mechanical integrity. This stress-dependent suppression of chain cleavage is therefore a direct mechanism by which the network can stabilize itself. Because these observations concern reactions occurring at the level of individual chains, rather than the macroscopic response of the network, we refer to this effect as microscopic catch bond behaviour. In the next section, the focus will be on the macroscopic response of the whole network instead, by looking at the strain rate. 

In Fig.~\ref{fig:FigResMicro}(b), the number of cleavage reactions is shown again, similar to panel (a), but now distinguishing between strictly local reactions and reactions without a locality constraint. As depicted in Fig.~\ref{fig:Fig2}, local refers to reactions that have taking place due to reactive groups located on the same chain. The solid lines in Fig.~\ref{fig:FigResMicro}(b) represent reactions that are strictly local, whereas the dotted lines were reactions that took place without a locality constrained, hence including both local and nonlocal reactions. The microscopic catch bond effect is clearly more pronounced when reactions are restricted to be local. This is consistent with the intuitive single-chain picture: increasing stress increases the average separation between reactive groups along a chain (Eq.~\ref{eq:Raffine}), thereby increasing the time required for them to encounter each other and react, as illustrated in Fig.~\ref{fig:Fig1}. When nonlocal reactions are allowed, this argument is less applicable, since reactive groups may be spatially close due to network topology rather than by opposing chain stretching. These results suggest that the intuitive single-chain mechanism is correct and that the observed catch bond behaviour predominantly originates from local reactions. This insight enables system tuning. For example, in an experimental setting, the chemistry could be designed to permit only local reactions, thereby enhancing the catch bond effect.

\begin{figure}[t!]
    \centering
    \includegraphics[width=\linewidth]{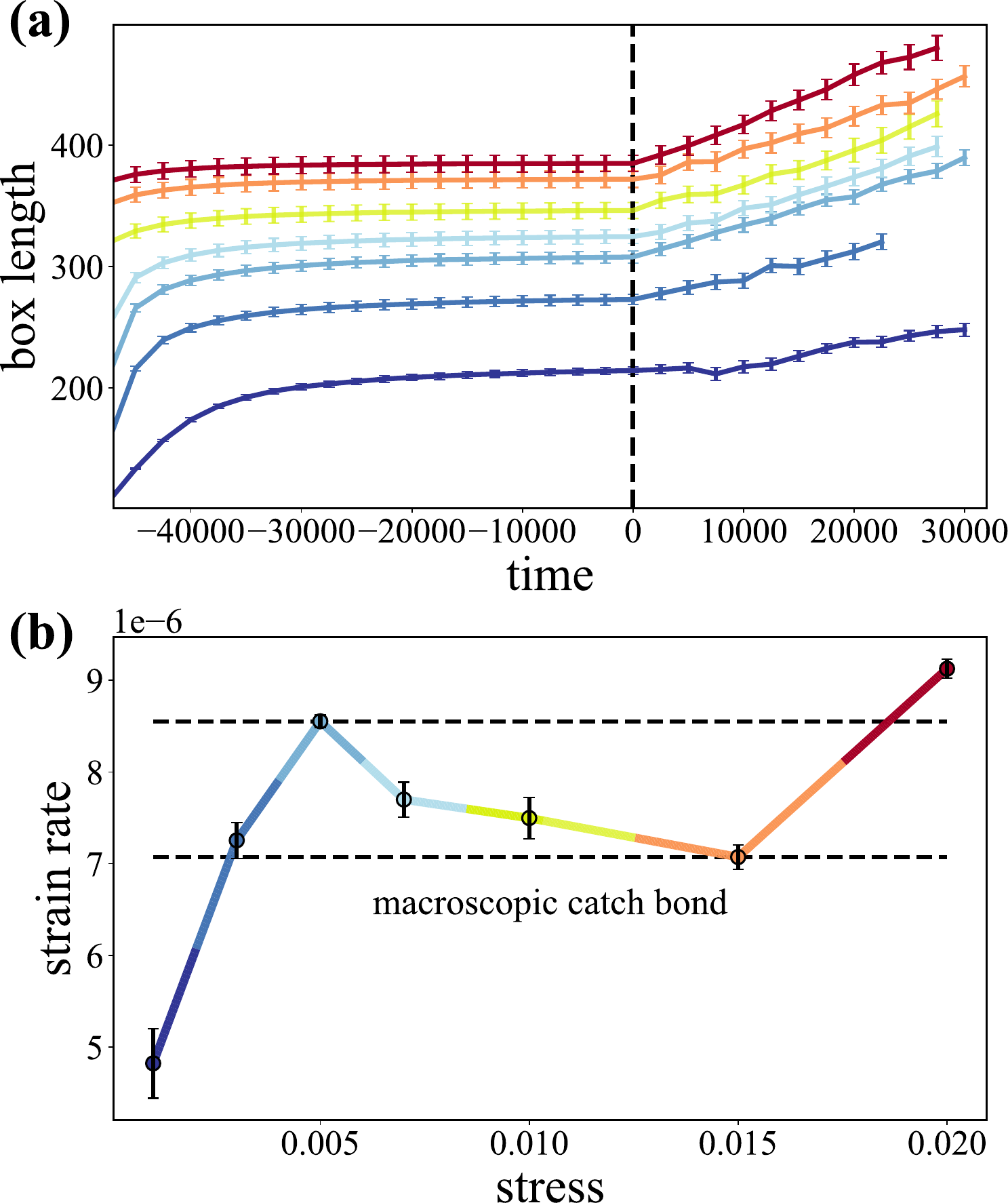}
    \caption{Macroscopic catch bond behaviour under applied uniaxial stress. (a) Evolution of the simulation box length along the loading direction during a tensile test for fixed target uniaxial stresses. The box length first equilibrates to a steady value while reactions are disabled and subsequently increases further once the reactions are reactivated, indicating additional deformation driven by chemical activity. (b) Strain rate as a function of the applied stress, obtained from the slope of the box length curves in panel (a) after reactions are reactivated at time $0$. A non-monotonic dependence of strain rate on applied stress is indicative of macroscopic catch bond behaviour, arising from stress-induced suppression of reactions that limits the rate of material elongation.}
    \label{fig:FigResMacro}
\end{figure}
In the Supplementary Material, we furthermore examine the number of ring-opening reactions and show that they are effectively limited by the number of ring-closing events, resulting in nearly identical reaction rates. This coupling ensures that the total number of rings in the network is largely conserved, preventing network degradation during tensile tests.

\subsection*{Non-monotonic strain rate}
Next, we examine the material's response to the applied stress, focusing on macroscopic deformation rather than individual reactions, during the same tensile tests. In Fig.~\ref{fig:FigResMacro}(a), we show the evolution of the simulation box length along the loading direction. As periodic boundary conditions were applied, this is directly related to the networks extension. During the first phase of the tensile test, the material deforms to a finite equilibration length while reactions are disabled, establishing a pre-stressed network. The elongation at time zero of this pre-stressed network with respect to the applied stress exhibits a non-linear behaviour, which is typical for polymer networks and well captured by the hyperelastic Ogden model~\cite{ogden1972large}, as we show in the Supplementary Material. Once reactions are reactivated at time zero, the network starts creeping further, indicating additional deformation driven by chemical activity. Importantly, in all simulations, the network remains intact throughout the tensile test.

The slope of the elongation curve corresponds to the strain rate. After reactivation of the reversible reactions, the strain rate was extracted and is shown in Fig.~\ref{fig:FigResMacro}(b). As expected for conventional viscoelastic materials, the lowest applied stress yields the smallest strain rate, while the highest stress produces the largest strain rate. Strikingly, this monotonic ordering breaks down at intermediate stress levels: increasing the applied stress can instead reduce the strain rate, such that the material extends slower under a higher applied stress. This results in a pronounced non-monotonic dependence of the strain rate on the applied stress. Such behaviour constitutes a macroscopic manifestation of catch bond mechanics. In this intermediate regime, higher stresses suppress cleavage reactions, thereby stabilizing load-bearing chains and limiting the rate of network rearrangement that enables deformation. As a consequence, the material stiffens dynamically under load, revealing how microscopic force-dependent reaction kinetics translate into emergent, stress-adaptive macroscopic mechanical response.

\section*{Discussion}
\label{sec:conc}
Catch bonds provide a valuable design principle for materials science due to their ability to strengthen under applied load~\cite{dansuk2023catch, rabbi2025fine}. However, few chemical systems simultaneously exhibit catch bond behaviour and lend themselves to a realistic synthetic design~\cite{dansuk2019simple, dansuk2021self, dansuk2023catch, van2020chemical, liu2024engineering}, slowing progress in this area. Here, we establish reversible ring formation as a general synthetic catch bond mechanism. Existing realizations of a reversible ring-forming chemistry, such as in covalently adaptable polymer networks~\cite{van2021reprocessing, delahaye2019internal}, confirm the experimental accessibility of this concept. More broadly, reversible ring formation itself emerges as a fundamentally general and versatile motif for engineering synthetic catch bond behaviour, opening up a wide design space that extends far beyond these specific network chemistries.

Through tensile tests on coarse-grained models of hydrogels incorporating the proposed reversible ring-forming chemistry, we find that increasing the applied stress suppresses reactions, despite a constant concentration of reactive groups, as was shown in Fig.~\ref{fig:FigResMicro}(a). This effect arises from stress-induced chain stretching, which increases the average separation between reactive groups on a chain and thereby prolongs the time required for a reaction to take place. This finding extends well-established single chain arguments~\cite{blumberg2005disruption, shin2012effects, laeremans2024polymer, laeremans2025theoretical} to crosslinked polymer networks. Because ring-closure transiently cleaves polymer strands and reduces network connectivity, stress-induced suppression of these reactions preserves load-bearing chains and enhances mechanical integrity, giving rise to catch bond behaviour. This effect is most pronounced when reactive groups are constrained to react locally along the same polymer chain, revealing a clear molecular design principle: enforcing reaction locality strongly amplifies catch bond performance, as depicted in Fig.~\ref{fig:FigResMicro}(b).

We next examined the macroscopic response of the network under tensile stress, focusing on overall deformation rather than individual reactions. In the simulations, the network initially creeps to a finite equilibrium length with reactions disabled, establishing a pre-stressed state. Once reversible reactions are reactivated, the network elongates further, driven by chemical activity. Remarkably, the resulting strain rate exhibits a pronounced non-monotonic dependence on applied stress: it increases at low and high stresses, but decreases at intermediate stresses, as shown in Fig~\ref{fig:FigResMacro}. This unconventional behaviour arises from the stress-dependent suppression of ring-closing reactions, which stabilizes load-bearing chains and slows network rearrangement. The effect demonstrates how microscopic catch bond kinetics translate into emergent, stress-adaptive macroscopic mechanics, providing a built-in self-regulation mechanism that could be exploited in responsive or impact-mitigating materials.

A particularly enticing class of responsive materials whose design could benefit from this type of catch bonds is that of trainable matter~\cite{jaeger2024training, boynton2024accessing}, materials that mimic the biological motif of network remodeling. Following the interpretations of Mulla and coworkers~\cite{mulla2022weak}, having catch bonds allows a material under load to reorganize its topology while the bonds that are currently contributing most to the load-bearing are more likely to remain in place, hence providing a route to make the materials more resilient to precisely the load that it is currently under.

In summary, we showed that reversible ring-forming polymers provide a versatile and intrinsically general route to stress-adaptive materials, as their catch bond behaviour is inherent to the chemistry and independent of specific implementations. This mechanism could be harnessed to create multifunctional hydrogels with built-in mechanical self-regulation, suitable for protective, impact-mitigating, or other load-responsive applications.


\subsection*{Methods}
\footnotesize \textbf{Molecular dynamics simulations.} All simulations were performed
using the molecular dynamics simulator \textsc{lammps}~\cite{LAMMPS}. The dimensionless Lennard-Jones unit system was used. The dynamics were simulated using Langevin dynamics at a temperature of 0.5. The timestep was taken to be 0.0005. Percolation in all three spatial dimensions was checked using \textsc{perconet}~\cite{perconet2022}. The number of rings was found using NetworkX~\cite{hagberg2008exploring}. Reactions were simulated using the \textsc{reacter}~\cite{gissinger2020reacter} package. In Fig.~\ref{fig:FigResMicro} and Fig.~\ref{fig:FigResMacro}, the results are averaged over the 90 longest runs per target stress. However, 5 runs in total were excluded because the stress exceeded a certain tolerance. Additional details are provided in the Supplementary Material. 

\textbf{Error estimation.} The default error used in this work is the standard error of the mean. Only for the result of the strain rate presented in Fig.~\ref{fig:FigResMacro}(b), a different modified error was used. To obtain this data, a linear least-squares fit was performed on the strains $\varepsilon(t)$ of Fig.~\ref{fig:FigResMacro}(a) after time $t$, constrained to pass through the origin:
\begin{align}
    \varepsilon(t)/\varepsilon(0) \approx a \, t,
\end{align}
where $a = d\varepsilon/dt$ denotes the slope, and hence represents the strain rate. A figure of the raw fits is provided in the Supplementary Material. The standard error of the slope is computed from the residuals $r_i = \varepsilon(t_i) - a\, t_i$ as
\begin{align}
    \text{SE}(a) = \sqrt{\frac{\sum_{i=1}^n r_i^2}{(n-1)\sum_{i=1}^n t_i^2}},
\end{align}
where $n$ is the number of data points. To account for deviations from perfect linearity, we define a coefficient of determination
\begin{align}
    R^2 = 1 - \frac{\sum_{i=1}^n r_i^2}{\sum_{i=1}^n (\varepsilon(t_i)-\bar{\varepsilon})^2}, \qquad
    \bar{\varepsilon} = \frac{1}{n}\sum_{i=1}^n \varepsilon(t_i),
\end{align}
with $0 < R^2 < 1$. The final combined error for the slope is then taken as
\begin{align}
    \sigma_a =
        \text{SE}(a)/\sqrt{R^2}.
\end{align}
This procedure ensures that the error bars reflect both statistical uncertainty and deviations from linear behaviour.

\textbf{Code availability.} Code will be made publicly available upon publication.

\textbf{Author contributions.}
\emph{WL} performed the simulations, analyzed the data, and wrote the manuscript. \emph{WGE} supervised the research process and provided manuscript revisions.

\textbf{Statement of competing interests.}
The authors declare that they do not have any competing interests.

\normalsize
\acknowledgments{This research is financially supported by the Dutch Ministry of Education, Culture and Science (Gravity Program 024.005.020 – Interactive Polymer Materials IPM). The authors are grateful for discussions with Roy Wink and Rint Sijbes\-ma.
}

\bibliography{references}

\section*{SUPPLEMENTARY INFORMATION}

\subsection*{Average polymer stretch under affine uniaxial deformation}
Denoting the initial end-to-end vector of a polymer as $\mathbf{R}$, an affine and volume preserving transformation of the form $\mathbf{F}(t) = \text{diag}(\lambda(t), 1/\sqrt{\lambda(t)}, 1/\sqrt{\lambda(t)})$ transforms the end-to-end vector as $\mathbf{R}' = \textbf{F}\textbf{R}$. Hence, the scalar $R'^2$ reads
\begin{align}
    R'^2 = \textbf{R}' \cdot \textbf{R}' = \left( \textbf{FR} \right)^\top \left( \textbf{FR} \right) = \textbf{R}^\top \cdot \textbf{F}^\top \textbf{F}\textbf{R}. \label{eq:gentrans}
\end{align}
Working out $\textbf{F}^\top \textbf{F}$ gives
\begin{align}
    \textbf{F}^\top \textbf{F} &= \begin{bmatrix}
\lambda^2 & 0 & 0\\
0 & 1/\lambda & 0 \\
0 & 0 & 1/\lambda
\end{bmatrix},
\end{align}
and writing $R^2 = R_x^2 + R_y^2 + R_z^2$, we find
\begin{align}
    R'^2 &= \lambda^2 R_x^2 + \frac{1}{\lambda} \left( R_y^2 + R_z^2 \right).
\end{align}
Next, defining the angle $\theta$ as the angle between the initial vector $\textbf{R}$ and the stretching direction $x$, we can use $R_x = \cos(\theta)R$. From this, it follows that $R_y^2 + R_z^2 = R^2 - R_x^2 = R^2 - \cos^2(\theta)R^2 = \sin^2(\theta)R^2$, by which we find
\begin{align}
    R'^2/R^2 = \lambda^2\cos^2(\theta) + 1/\lambda \sin^2(\theta).
\end{align}
Assuming that all chain orientations are homogeneous before stretching, we can average over the solid angle: $\langle \cos^2(\theta) \rangle = 1/3$ and $\langle \sin^2(\theta) \rangle = 2/3$, which yields~\cite{rubinstein}
\begin{align}
    \langle R'^2 \rangle &= \frac{\lambda^2 + 2/\lambda}{3} \, R^2, 
\end{align}
as was given in the main text in Eq.~5.

\subsection*{Average polymer stretch under any affine transformation}
Next, we show that for any affine and volume preserving transformation $\textbf{F}$, the property $\langle R'^2 \rangle \geq R^2$ holds. We start from the general transformation (Eq.~\ref{eq:gentrans})
\begin{align}
    R'^2 = \textbf{R}^\top \cdot \textbf{F}^\top \textbf{F}\textbf{R}. 
\end{align}
Given that the orientations of the polymers are initially homogeneous (before the deformation), we can use that $\langle R_i R_j \rangle = \delta_{ij}\frac{R^2}{3}$, with $i,j \in \{ x,y,z \}$. Hence, we find the expression
\begin{align}
    \langle R'^2 \rangle &= \frac{R^2}{3} \text{Tr}(\textbf{F}^\top \textbf{F)}. \label{eq:tracetrace}
\end{align}
Furthermore, calling $\mu_1, \mu_2, \mu_3$ the eigenvalues of $\textbf{F}^\top \textbf{F}$ we have
\begin{align}
    \text{det}(\textbf{F}^\top \textbf{F}) \nonumber &= \mu_1\mu_2\mu_3 \\ &= \text{det}(\textbf{F})^2 = 1, \label{eq:isone} 
\end{align}
where we used the property of square matrices $\text{det}(\textbf{F}^\top \textbf{F}) = \text{det}(\textbf{F}^\top) \text{det}(\textbf{F})$ and the fact that $\text{det}(\textbf{F}^\top) = \text{det}(\textbf{F})$. Furthermore, we can use that
\begin{align}
    \text{Tr}(\textbf{F}^\top \textbf{F}) &=  \mu_1 + \mu_2 + \mu_3.
\end{align}
Knowing the aritmic mean - geometric mean inequality
\begin{align}
    \frac{\mu_1 + \mu_2 + \mu_3}{3} \geq \left( \mu_1\mu_2\mu_3 \right)^{1/3},
\end{align}
and knowing that $\mu_1\mu_2\mu_3 = 1$ (Eq.~\ref{eq:isone}), we have
\begin{align}
    \text{Tr}(\textbf{F}^\top \textbf{F}) \geq 3.
\end{align}
Combining this with the expression in Eq.~\ref{eq:tracetrace}, we have proven that any affine, volume preserving transformation leads to
\begin{align}
    \langle R'^2 \rangle \geq R^2.
\end{align}
Hence, it is expected that catch bond behaviour would also be observed in biaxial or shear deformation experiments. 

\subsection*{Implementation of reactions}
Reactions were implemented using the \textsc{reacter}~\cite{gissinger2020reacter} package. To maintain numerical stability, reactive groups that had just participated in a reaction were excluded from further reactions for 200 timesteps. At each timestep, the network topology was examined for reactive groups satisfying the local criterion for a ring-closing reaction. Specifically, groups were considered eligible if they were located on the same polymer chain, separated by exactly six beads, and within a reaction distance of $\sigma \times 2^{1/6}$. When these conditions were met, the reaction was executed with a probability of 0.001. After each reaction attempt, the local criterion was reevaluated, and reactive groups that still satisfied the condition but did not react were again excluded from further reactions for 200 timesteps, also with probability 0.001. This lowered the reaction rate more to preserve network integrity. The same protocol was then applied without the local constraint, allowing nonlocal ring-closing reactions. Finally, the network topology was checked for ring-opening reactions, which were executed with unit probability when reactive groups were within a distance of $\sigma$.

\begin{figure}[t!]
    \centering
    \includegraphics[width=\linewidth]{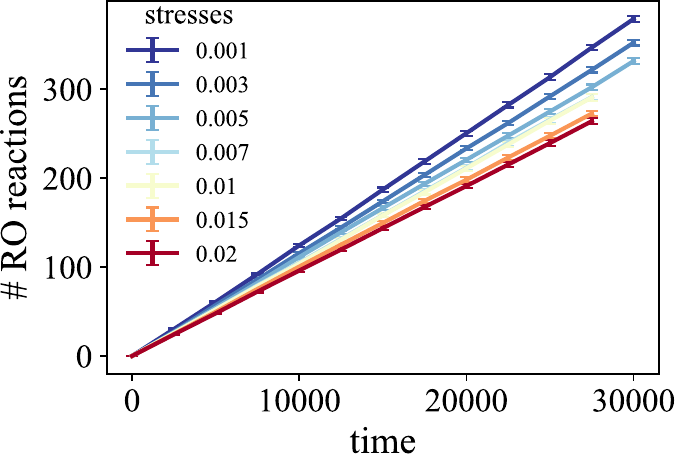}
    \caption{Number of ring-opening (RO) reactions as a function of time after reactivation of the reversible RO reaction, for fixed target uniaxial stresses of increasing magnitude.}
    \label{fig:FigRO}
\end{figure}

The use of small reaction probabilities and restricted reaction distances for ring-closing reactions was chosen to preserve network integrity throughout the simulations. Nevertheless, the qualitative behaviour reported is expected to depend only weakely on the exact values.

\subsection*{Number of ring-opening reactions}
In Fig.~5 of the main text, we showed that the number of ring-closing reactions decreases with increasing target stress. This reaction is the primary contributor to the catch bond effect, as it effectively cleaves polymer chains. Nevertheless, the ring-opening reaction, which reconnects chains, was made highly probable by choosing a large capture radius and a reaction probability of unity. As a result, the ring-opening and ring-closing reactions are intrinsically coupled. Although the ring-opening reaction is significantly faster, it can only occur once rings are present. Consequently, the number of ring-opening events closely matches the number of ring-closing events, leading to a similar lowering in quantity with increasing target stress, as shown in Fig.~\ref{fig:FigRO}.

\begin{figure}[t!]
    \centering
    \includegraphics[width=\linewidth]{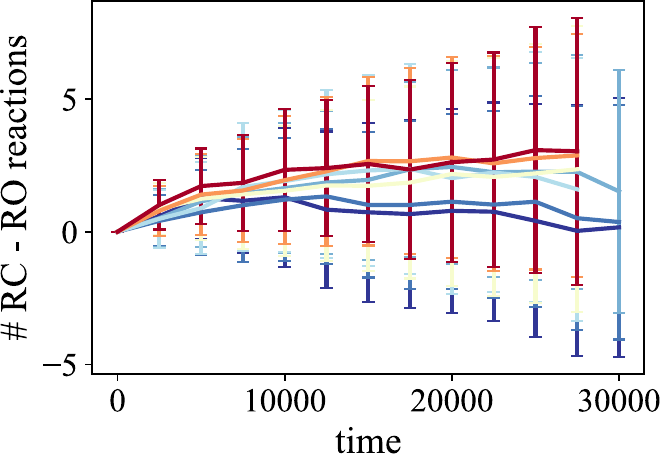}
    \caption{Number of ring-closing (RC) minus ring-opening (RO) reactions as a function of time after reactivation of the reversible RO reaction, for fixed target uniaxial stresses of increasing magnitude.}
    \label{fig:FigRCminRO}
\end{figure}

This, in turn, implies that even though a tensile test is performed, the network does not degrade. By examining the difference between the number of ring-closing (RC) and ring-opening (RO) reactions (Fig.~\ref{fig:FigRCminRO}), which corresponds to the number of chains broken in excess of those present prior to the tensile test, we find that this quantity does not deviate significantly from zero. Hence, the total number of rings is effectively conserved. Moreover, a slight indication of network healing is observed at increasing stress, although this effect is not statistically significant.

\subsection*{Target stress protocol and excluded runs}
\begin{figure}[t!]
    \centering
    \includegraphics[width=\linewidth]{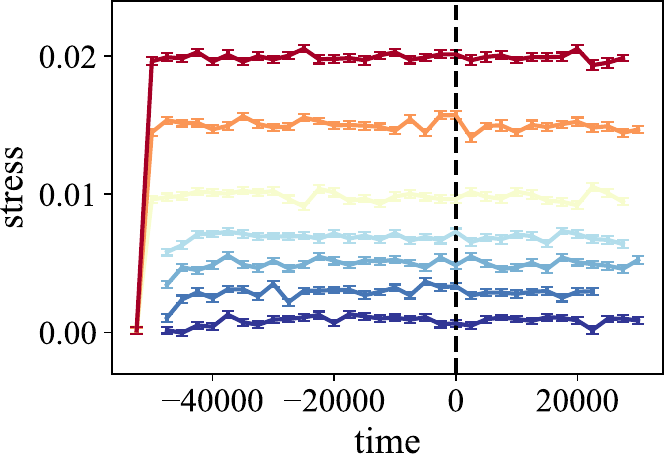}
    \caption{Observed (true) stress over time for increasing target stress for the 90 longest runs. When after time -40 000, \(|\text{stress} - \text{target stress}| \geq 0.02\), the runs were excluded, which was the case for 3 runs at target stress 0.01 and 5 runs at target stress 0.02.}
    \label{fig:TargetStress}
\end{figure}

The target stress was applied using the \textit{fix deform/pressure} command in \textsc{lammps}~\cite{LAMMPS}. For the analysis, the 90 longest simulation runs were selected. An additional filtering step was then applied: runs for which, after a time of \(-40\,000\), the deviation \(|\text{stress} - \text{target stress}| \geq 0.02\) were excluded. Such deviations arise from the finite size of the simulation box and would likely not occur in larger systems or in real experiments. Nevertheless, due to computational limitations, a small number of such cases were present and therefore excluded from the analysis. Specifically, this affected three runs at a target stress of 0.01 and five runs at a target stress of 0.02. As shown in Fig.~\ref{fig:TargetStress}, the measured stress rapidly relaxes to the target value and remains nearly constant after the reversible reactions are activated at time \(t = 0\).

\begin{figure}[b!]
    \centering
    \includegraphics[width=\linewidth]{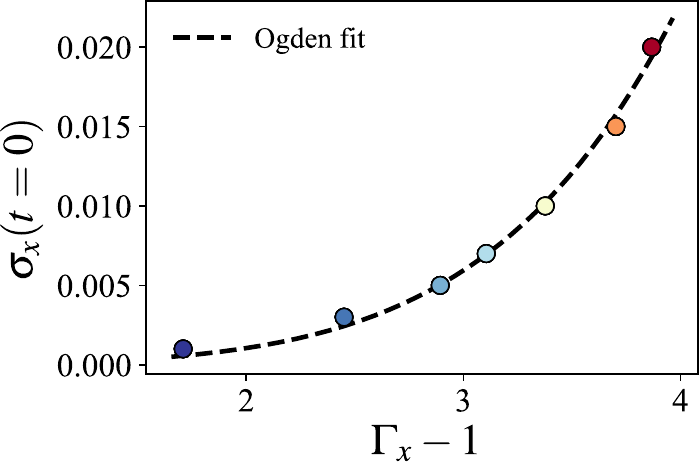}
    \caption{Stress of pre-stressed material $\sigma_x(0)$ against $\Gamma_x - 1 = (\ell_x(0) - \ell_x(-\infty))/\ell_x(-\infty)$. The model follows a non-linear behaviour, typical for polymer networks under high stress. A fit was performed using the Ogden model (Eq.~\ref{eq:Ogden}), which resulted in $\mu = 1.396 \times 10^{-6} \pm 6.400 \times 10^{-7}$ and $\alpha = 6.027 \pm 0.298$.}
    \label{fig:OgdenFit}
\end{figure}
\subsection*{Non-linear stress-strain relation of pre-stessed material}
In Fig.~6(a) of the main text, the extension of the network along the loading direction was plotted for increasing stress. For the pre-stressed material, one can observe a non-linear relation. That is, at time zero, right before the reactions are turned on again, the material seems to have extended in a non-linear way with regard to the target stress. This is not surprising for polymer networks, and can be explained using the Ogden hyperelastic model~\cite{ogden1972large}. This model expresses the stress of the pre-stresses material $\sigma_x(t = 0)$ in function of the relative extension $\Gamma_x \equiv \ell_x(t=0)/\ell_x(t \rightarrow -\infty)$ under uniaxial tension as
\begin{align}
    \sigma_x(t=0) &= \mu \left( \Gamma_x^\alpha - \Gamma_x^{-\alpha/2}\right), \label{eq:Ogden}
\end{align}
with $\mu$ and $\alpha$ material properties.

In Fig.~\ref{fig:OgdenFit}, $\sigma_x$ is plotted against $(\ell_x(0) - \ell_x(-\infty))/\ell_x(-\infty) = \Gamma_x - 1$. Here, it can be seen that the Ogden model indeed describes the data well. From the fit, we obtained $\mu = 1.396 \times 10^{-6} \pm 6.400 \times 10^{-7}$ and $\alpha = 6.027 \pm 0.298$.

\begin{figure}[b!]
    \centering
    \includegraphics[width=\linewidth]{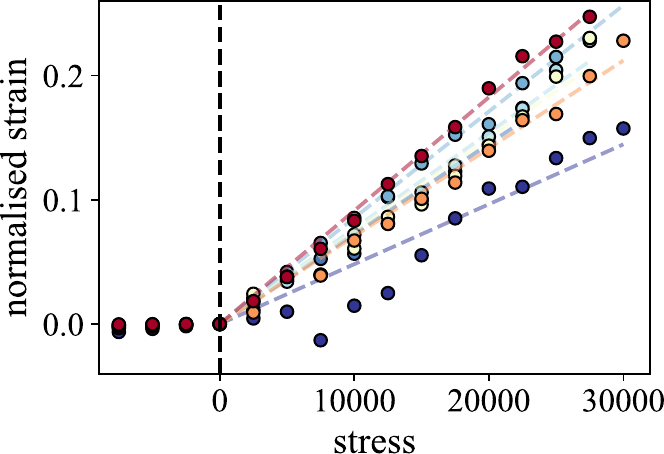}
    \caption{Normalized strain vs stress after time zero, fitted with a linear line used for the result of Fig.~6(b). }
    \label{fig:fities}
\end{figure}

\subsection*{Fits for strain rate}
In Fig.~\ref{fig:fities}, the linear fits of the strain after time zero are shown. These were used for the strain rate as reported in the main text in Fig.~6(b). Most pronounced for the very lowest stress, some deviation from a linear line can be observed. Hence, deviation from linearity was included in the error bars, as explained in the Methods section.

\end{document}